\documentclass[twocolumn]{aastex631}
\usepackage{graphicx} 

\shorttitle{Planetary Habitability Under a Changing Star}
\shortauthors{Fetherolf et al.}

\begin{document}

\title{Planetary Habitability Under the Light of a Rapidly Changing Star}

\author[0000-0002-3551-279X]{Tara Fetherolf}
\altaffiliation{NASA Postdoctoral Program Fellow}
\affiliation{Department of Earth and Planetary Sciences, University of California, Riverside, CA 92521, USA}
\email{tara.fetherolf@gmail.com}

\author[0009-0000-4164-2358]{Sadie G. Welter}
\affiliation{Department of Earth and Planetary Sciences, University of California, Riverside, CA 92521, USA}

\author[0000-0002-7084-0529]{Colby M. Ostberg}
\affiliation{Department of Earth and Planetary Sciences, University of California, Riverside, CA 92521, USA}
\affiliation{Laboratory for Atmospheric and Space Physics, University of Colorado, 600 UCB, Boulder, CO 80309, USA}

\author[0000-0002-7084-0529]{Stephen R. Kane}
\affiliation{Department of Earth and Planetary Sciences, University of California, Riverside, CA 92521, USA}

\author[0000-0001-6487-5445]{Rory Barnes}
\affiliation{Department of Astronomy and Astrobiology Program, University of Washington, Box 351580, Seattle, WA 98195, USA}
\affiliation{NExSS Virtual Planetary Laboratory, Box 351580, University of Washington, Seattle, WA 98195, USA}

\author[0000-0003-0447-9867]{Emilie R. Simpson}
\affiliation{SETI Institute, Mountain View, CA  94043, USA}

%%%%%%%%%%%%%%%%%%%%%%%%%%%%%%%%%%%%%%%%%%%%%%%%%%%%%%%%%%%%%%%%%%%%

\begin{abstract}

Planetary atmospheric energy budgets primarily depend on stellar incident flux. However, stellar variability can have major consequences for the evolution of planetary climates. In this work, we evaluate how stellar variability influences the equilibrium temperature and water retention of planets within the Habitable Zone (HZ). We present a sample of 9 stars that are known to host at least one planet within the HZ and that were identified to have a variability amplitude exceeding 100~ppm based on photometry from the Transiting Exoplanet Survey Satellite (TESS). We investigate the effect that the variability of these stars have on the insolation flux of their HZ planets and the resulting changes in the induced planetary equilibrium temperature. Our results show that for the stars in our sample, the stellar variability has an insignificant effect on the equilibrium temperature of HZ planets. However, we also emphasize that these stars are not representative of more extreme variable stars, since exoplanets are more difficult to detect and characterize in the presence of extreme variability. We also investigate the equilibrium temperature and long-term evolution of a hypothetical Earth-like planet placed at the inner edge of the HZ around a highly variable star. We found that the water loss rates are comparable between both variable and quiet host stars for Earth-like planets in the inner HZ. Overall, these results broaden our knowledge on the impact of stellar variability on planetary habitability.

\end{abstract}

%%%%%%%%%%%%%%%%%%%%%%%%%%%%%%%%%%%%%%%%%%%%%%%%%%%%%%%%%%%%%%%%%%%%

\section{Introduction}
\label{sec:intro}

As the dominant contribution to a planetary atmospheric energy budget, the insolation flux received from a star has a large influence on the habitability of a planet. The energy received by a planet from its star also has major consequences for Habitable Zone (HZ) distances, atmospheric chemistry, and planetary surface temperatures. Thus, periodic changes in the flux received by a planet can have a significant effect on the evolution of the planetary climate. Stellar variability can arise from numerous different sources, including sunspots, pulsations, and rotation \citep[e.g.,][]{pojmanski2002,wright2011d,eyer2019,fetherolf2023b}. With the improvement in the precision of photometric measurements provided by observatories and space telescopes, the fraction of stars showing variations in their luminosity is increasing, furthering the need to study this aspect of stellar astrophysics \citep{ciardi2011,basri2011,eyer2019,briegal2022,fetherolf2023b}. Primary data sources for precision photometry include those from the Kepler mission \citep{borucki2010a} and the Transiting Exoplanet Survey Satellite \citep[TESS;][]{ricker2015}. Such data have enabled significant progress towards understanding the variability of stars across the sky compared to the Sun, where it has been found that a large fraction of stars are more variable than our Sun \citep{mcquillan2012,basri2013,walkowicz2013b}. Furthermore, using the precision of the original data release of Kepler \citet{ciardi2011} found that $>95$\% of G, K, and M stars exhibit variability.

The relatively large number of stars that are variable provides the research community with a plethora of opportunities to characterize the environments around variable stars \citep{howell2016a}, particularly those already known to host exoplanets \citep{kane2021b,simpson2023}. Recently, \citet{fetherolf2023b} presented more than 84,000 variable stars across the entire sky in the TESS Stellar Variability Catalog (TESS-SVC)---including $\sim$65,000 newly identified variable stars---using just the first two years of data from the TESS Prime Mission. Photometric monitoring of known host stars has previously been used to study how their variability impacts planetary signatures, such as the cases of HD~63454 \citep{kane2011e} and HD~192263 \citep{henry2002,dragomir2012a}. Photometric variability has also been attributed to false-positive exoplanet detections \citep{Robertson14,Robertson14-1,Robertson15,sullivan2015,kane2016a,Hojjatpanah20,Prajwal22}. \citet{simpson2022a}, for example, identified BD-06~1339~b as a false positive planet based on a strong correlation between the radial velocity (RV) data and the star's photometric variability. Given that this is not the first instance of photometric variability being able to confirm or refute the detection of a planet, it is clear that understanding stellar variability is important for fully characterizing exoplanetary systems. 

In our own Solar System, stellar variability presents itself over the course of the Sun's 11 year cycle. It is well known that dark sunspots that appear on our Sun cause variations in its luminosity \citep{babcock1961,leighton1969,hathaway2015}. Despite observable data showing that the Sun's total solar irradiance decreases as a sunspot forms, \citet{foukal2006} showed that this did not have a significant impact on the Earth’s climate evolution over time. However, it has not yet been fully established at what level of stellar variability that would induce a non-negligible effect on the climate evolution of a star's planets, and thus there could be potential for significantly variable stars to have an impact on the potential habitability of their worlds. \citet{simpson2023} analyzed 263 variable stars with 337 planetary companions and found an observational bias that favors the detection of giant planets around variable stars. This bias presents an opportunity to also study the case of exomoons around giant planets, whose incident flux profile becomes increasingly complex when accounting for both their orbit around the giant planet and the variability of the host star \citep{heller2013a,hinkel2013b}. Such studies regarding the variability of incident flux on terrestrial bodies may be applied to a variety of scenarios that consider the robustness of planetary climates and potentially expand the possible sustainability of habitable conditions \citep{wolf2019}. Clearly, the prevalence of variable stars, their relative lack of representation in exoplanet discoveries, and their influence on planetary climates requires further analysis of their overall system properties.

In this work, we investigate how stellar variability influences the planetary energy balance and water loss. Section~\ref{sec:sample} describes the methodology for acquiring a list of 9 known HZ exoplanets with variable host stars. The properties of the stars and planets in our sample are reviewed in Section~\ref{sec:sample_stats}. We present the fluxes received by the planets and their resulting equilibrium temperatures, further discuss the findings for 3 of the 9 targets in our sample, and summarize how stellar variability influences the climates of these HZ exoplanets in Section~\ref{sec:teq}. We then explore the conditions of a hypothetical Earth-like planet in the HZ of an extreme variable star in Section~\ref{sec:vplanet}, including modeled results of water loss that occurs on the planet over the lifetime of the host star. We discuss the implications for exomoons in Section~\ref{sec:exomoons}, and lastly state our concluding remarks in Section~\ref{sec:conclusions}.

%%%%%%%%%%%%%%%%%%%%%%%%%%%%%%%%%%%%%%%%%%%%%%%%%%%%%%%%%%%%%%%%%%%%

\section{Sample Selection} 
\label{sec:sample}

Our initial sample of variable stars is collected from the TESS Stellar Variability Catalog \citep[TESS-SVC;][]{fetherolf2023b}. The TESS-SVC is publicly available as a High-Level Science Product on the Mikulski Archive for Space Telescopes\footnote{\url{https://archive.stsci.edu/}} (MAST) and includes a total of 84,046 variable stars that were observed at a 2-min cadence during the TESS Prime Mission (\dataset[DOI: 10.17909/f8pz-vj63]{http://dx.doi.org/10.17909/f8pz-vj63}). From this sample of variable stars, \citet{simpson2023} identified 263 variable stars that harbor at least one confirmed exoplanet, which include a total of 337 planets. We then cross matched the list of known exoplanets around variable stars from \citet{simpson2023} with planets that exist in the Habitable Zone (HZ) of their host stars \citep{kasting1993a,kane2012a,kopparapu2013a,kopparapu2014,kane2016c,hill2018,hill2023} based on the data available in the Habitable Zone Gallery\footnote{\url{http://www.hzgallery.org/}}. Planets that spend less than 50\% of their orbit within the optimistic HZ (OHZ) were removed, as well as any planets that are in circumbinary systems, which resulted in a sample of 16 confirmed HZ exoplanets around variable stars. 

The TESS-SVC variability analysis \citep{fetherolf2023b} was only based on an single sector of TESS photometry. We choose to perform a follow-up variability analysis on all available sectors from the TESS Prime Mission (sectors 1--26) for these 16 systems. We use the Presearch Data Conditioning Simple Aperture Photometry \citep[PDCSAP;][]{Stumpe12, Stumpe14, Smith12} 2-min light curves that were processed by the Science Processing Operations Center (SPOC) pipeline \citep{Jenkins16} and are available on MAST (\dataset[DOI: 10.17909/t9-nmc8-f686]{http://dx.doi.org/10.17909/t9-nmc8-f686}). We follow a similar methodology to \citet{fetherolf2023b}, but apply a median normalization for each sector and perform a Lomb-Scargle periodogram search \citep{lomb76,scargle82} in the range of 1 day to half of the observed baseline. For our final sample, we require the amplitude of measured variability to be at least 100\,ppm, which resulted in 9 HZ exoplanets around variable stars.

\begin{deluxetable*}{lcccccc}
  \tablewidth{0pc}
  \tablecaption{\label{tab:sample} Stellar and orbital properties of 9 known HZ planets around variable stars.}
  \tablehead{
\colhead{Planet} & \colhead{TIC ID} & \colhead{$T_\mathrm{eff}$\tablenotemark{a}} & \colhead{$A$} & \colhead{$P_\mathrm{var}$} & \colhead{$P_\mathrm{orb}$\tablenotemark{a}} & \colhead{$e$\tablenotemark{a}} \\ [-0.2cm]
\colhead{} & \colhead{} & \colhead{(K)} & \colhead{(ppm)} & \colhead{(days)} & \colhead{(days)} & \colhead{}
  }
  \startdata
    TOI-1227 b & 360156606 & $3055\pm157$ & $7687\pm184$ & $1.662\pm0.019$ & $27.364\pm0.0001$ & $0$ \\
    HD 142415 b & 342041655 & $5943\pm135$ & $1356\pm3$ & $6.058\pm0.564$ & $386.3\pm1.6$ & $0.5$ \\
 &  &  & $711\pm3$ & $10.318\pm1.462$ &  &  \\
    HD 147513 b & 350673608 & $5873\pm106$ & $1014\pm1$ & $6.099\pm0.587$ & $528.4\pm6.3$ & $0.26\pm0.05$ \\
 &  &  & $620\pm1$ & $10.551\pm1.471$ &  &  \\
    HD 221287 b & 231716612 & $6376\pm100$ & $415\pm3$ & $5.764\pm0.482$ & $456.1^{+7.7}_{-5.8}$ & $0.08^{+0.17}_{-0.05}$ \\
 &  &  & $352\pm3$ & $2.678\pm0.095$ &  &  \\
    BD-08 2823 c & 33355302 & $4816\pm107$ & $336\pm7$ & $6.354\pm0.535$ & $237.6\pm1.5$ & $0.19\pm0.09$ \\
    KELT-6 c & 165987272 & $6266\pm105$ & $207\pm16$ & $7.660\pm0.812$ & $1276^{+81}_{-67}$ & $0.21^{+0.039}_{-0.036}$ \\
    HD 238914 b & 359558085 & $4769\pm15$ & $164\pm2$ & $2.743\pm0.040$ & $4100\pm210$ & $0.56^{+0.07}_{-0.05}$ \\
    HD 147379 b & 230073581 & $3923\pm157$ & $130\pm1$ & $10.523\pm0.152$ & $86.78^{+0.16}_{-0.15}$ & $0.063^{+0.047}_{-0.038}$ \\
    HD 63765 b & 340697083 & $5439\pm136$ & $124\pm2$ & $10.724\pm0.483$ & $358\pm1$ & $0.24\pm0.04$
  \enddata
  \tablecomments{HD~142415, HD~147513, and HD~221287 show double-sinusoidal variability, and thus each have two variability amplitudes and two variability periods.}
  \tablenotetext{a}{Obtained from the NASA Exoplanet Archive.}
\end{deluxetable*}

%%%%%%%%%%%%%%%%%%%%%%%%%%%%%%%%%%%%%%%%%%%%%%%%%%%%%%%%%%%%%%%%%%%%%%%%%%

\section{Habitable Zone Exoplanets with Variable Host Stars}
\label{sec:sample_stats}

Stellar and planetary properties of the 9 targets in our sample are obtained from the NASA Exoplanet Archive \citep{nea_ref}, which is accessiable via \dataset[DOI: 10.26133/NEA12]{http://dx.doi.org/10.26133/NEA12}. The stellar effective temperature, stellar variability amplitude, stellar variability period, orbital period of the planet, eccentricity of the planetary orbit, and associated uncertainties are provided for the 9 targets in Table~\ref{tab:sample}. Our sample includes FGKM dwarf stars with stellar masses ranging from 0.17--1.25~$M_\odot$, and one giant star (HD~238914) with a mass of 1.47~$M_\odot$. The amplitudes of their variability range from $\sim$100--8000~ppm and periods of variability range from $\sim$1.6--10.7~days. HD~147379~b and HD~147513~b are in widely separated ($>$200\,AU) binary systems, and are both assumed to be around the primary stars in their systems. 

\begin{figure}
    \centering
    \includegraphics[width=\linewidth]{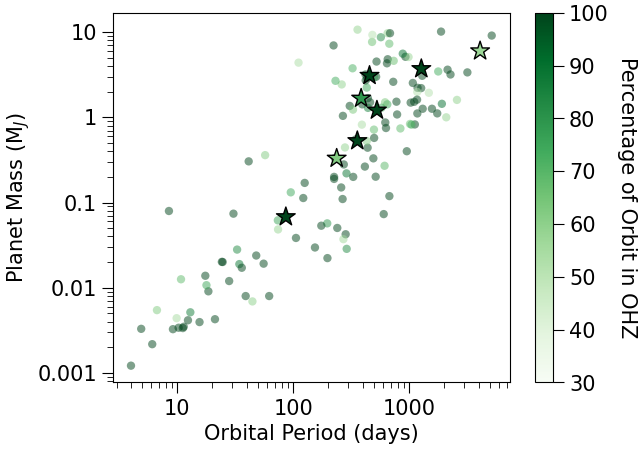}
  \caption{Orbital period versus planet mass for exoplanets in the Habitable Zone Gallary that spend greater than 50\% of their orbit within the OHZ. The points are colored by the time spent in the OHZ. The star-shaped points represent 8 of the 9 planetary systems in our sample. TOI-1227~b does not have a tabulated mass in the NASA Exoplanet Archive and thus does not appear in this plot, but is otherwise reported to have an upper limit mass of $<1.7M_J$ \citep{Mann22}.}
  \label{fig:sample}
\end{figure}

Figure~\ref{fig:sample} shows the orbital period, planet mass, and percentage of time in the OHZ for the planets in our sample (except TOI-1227~b) relative to other HZ planets. TOI-1227~b is not shown because it does not have a tabulated planet mass in the NASA Exoplanet Archive, but it is otherwise reported to have an upper limit mass of $<1.7M_J$ \citep{Mann22}. While the average orbital period and planet mass for planets in our sample (828~days and 2.1~$M_J$, respectively) are generally larger than that of the broader sample of HZ planets (456~days and 1.7~$M_J$, respectively), the two samples are statistically consistent with being drawn from the same parent sample according to a two-sided t-test. Similarly, six of the 9 planets in our sample spend 100\% of their orbit in the OHZ, which is consistent with the 164 HZ planets that are found to spend 100\% of their time in the OHZ compared to the larger sample of 255 HZ planets spending $>$50\% of their orbit in the OHZ. HD~142415~b, BD-08~2823~c, and HD~238914~b spend less than 100\% of their orbit in the OHZ due to their higher eccentricities and average orbital distance from their host star. 

\begin{figure}
    \centering
    \includegraphics[width=\linewidth]{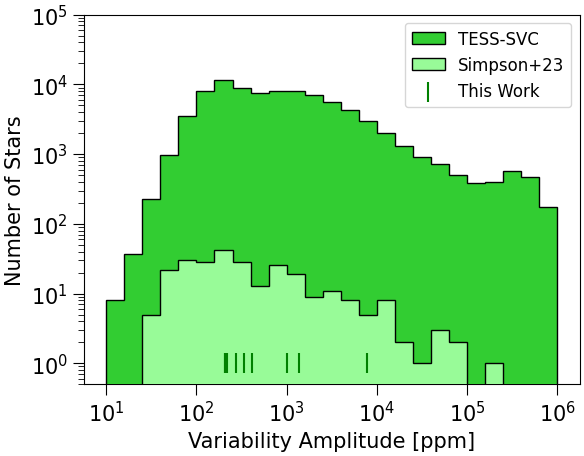}
  \caption{Histogram of the stellar variability amplitudes reported in the TESS-SVC \citep[dark green region,][]{fetherolf2023b}, the known exoplanet host stars \citep[light green region,][]{simpson2023}, and the 9 stars in our sample (dark green vertical lines) that all have at least one exoplanet that spends $>$50\% of its orbit in the OHZ.}
  \label{fig:var_amps}
\end{figure}

Figure~\ref{fig:var_amps} shows how the stellar variability amplitudes of the 9 stars in our sample (dark green vertical lines) compare to those in the TESS-SVC \citep[dark green region,][]{fetherolf2023b} and the variable stars with known exoplanets \citep[light green region,][]{simpson2023}. The average and median stellar variability amplitude of stars in the TESS-SVC is 11,350~ppm and 680~ppm, respectively. None of the stars in our sample exceed the average amplitude of the stars in the TESS-SVC, and only 3 stars exceed the median variability amplitude. Approximately 11\% of stars in the TESS-SVC have variability amplitudes higher than the star with the highest variability amplitude in our sample (TOI-1227), and 36\% of stars in the TESS-SVC have variability amplitudes higher than the next highest variability amplitude star in our sample (HD~142415). The known exoplanet variable host stars, on the other hand, have much lower average and median variability amplitudes of 3140~ppm and 260~ppm, respectively. This means that one star in our sample exceeds the average amplitude of variable exoplanet host stars and five stars exceed the median variability amplitude. A total of 8\% of the known variable exoplanet host stars have amplitudes higher than TOI-1227, and 23\% of the known variable exoplanet host stars have variability amplitudes higher than HD~142415. 

%%%%%%%%%%%%%%%%%%%%%%%%%%%%%%%%%%%%%%%%%%%%%%%%%%%%%%%%%%%%%%%%%%%%%%%%%%

\section{Variability Effects on Planet Equilibrium Temperatures}
\label{sec:teq}

In order to investigate how stellar variability may affect HZ exoplanets, we calculate the change in flux received by the planets in our sample and the estimated equilibrium temperature of the planets throughout their orbit. In order to estimate fluxes, we use the orbital period, semi-major axis, and stellar luminosity available on the NASA Exoplanet Archive \citep[][see Table~\ref{tab:sample}]{nea_ref}. If the luminosity of the star was not directly available, we calculated the luminosity using the stellar radius and effective temperature reported in the NASA Exoplanet Archive. We incorporate the change in flux attributed to stellar variability by using the variability amplitudes and periods that were measured from the variability analysis described in Section~\ref{sec:sample}. For most stars, the variability is consistent with being represented by a single sinusoidal function, but HD~142415~b, HD~147513~b, and HD~221287~b are best represented by the sum of two sinusoidal functions (see Table~\ref{tab:sample}). The minimum and maximum fluxes received by the planets in our sample, attributed separately to the orbital eccentricity and the stellar variability, are listed in Table~\ref{tab:teq}.

Equilibrium temperatures of the planets in our sample are estimated by assuming a 0 albedo and full heat redistribution ($f=1/4$). For eccentric systems, we choose to estimate temperatures when the planet is at periastron. The minimum and maximum estimated equilibrium temperatures of the planets in our sample, calculated separately for changes due to the orbit eccentricity and the stellar variability, are reported in Table~\ref{tab:teq}. Two decimal points are provided in order to to properly show very small estimated changes in equilibrium temperature. 

\begin{deluxetable}{lcccc}
    \tablewidth{0pc}
    \tablecaption{\label{tab:teq} Calculated values for the minimum and maximum flux received by the planet and its equilibrium temperature based separately on its orbital eccentricity and stellar variability.}
    \tablehead{
        \colhead{Planet} & \colhead{$F_{p,min}$} & \colhead{$F_{p,max}$} & \colhead{$T_{eq,min}$} & \colhead{$T_{eq,max}$} \\ [-0.2cm]
        \colhead{} & \colhead{(W/m$^2$)} & \colhead{(W/m$^2$)} & \colhead{(K)} & \colhead{(K)} 
    }
    \startdata
        \multicolumn{5}{l}{\bf{Orbit Eccentricity}} \\ [0.1cm] 
        Hypo Earth & 1479 & 1479 & 455.99 & 455.99 \\
        TOI-1227 b & 435 & 435 & 209.33 & 209.33 \\
        HD 142415 b & 653 & 5878 & 231.65 & 401.23 \\
        HD 147513 b & 480 & 1394 & 214.58 & 280.00 \\
        HD 221287 b & 1239 & 1708 & 271.89 & 294.58 \\
        BD-08 2823 c & 455 & 982 & 211.64 & 256.52 \\
        KELT-6 c & 531 & 1246 & 219.98 & 272.25 \\
        HD 238914 b & 1218 & 15319 & 270.74 & 509.79 \\
        HD 147379 b & 1230 & 1584 & 271.42 & 289.09 \\
        HD 63765 b & 548 & 1459 & 221.72 & 283.21 \\
        \hline
        \\ [-0.4cm] \multicolumn{5}{l}{\bf{Stellar Variability}} \\ [0.1cm] 
        Hypo Earth & 1452 & 1505 & 453.93 & 458.01 \\
        TOI-1227 b & 432 & 438 & 208.93 & 209.74 \\
        HD 142415 b & 5866 & 5890 & 401.02 & 401.44 \\
        HD 147513 b & 1391 & 1396 & 279.88 & 280.11 \\
        HD 221287 b & 1706 & 1709 & 294.53 & 294.64 \\
        BD-08 2823 c & 981 & 982 & 256.50 & 256.54 \\
        KELT-6 c & 1245 & 1246 & 272.24 & 272.27 \\
        HD 238914 b & 15316 & 15321 & 509.77 & 509.81 \\
        HD 147379 b & 1584 & 1584 & 289.08 & 289.10 \\
        HD 63765 b & 1459 & 1459 & 283.21 & 283.22
    \enddata
    \tablecomments{Hypo Earth is based on the G-type variable star TIC~80127634 with an hypothetical Earth-like planet in a circular orbit at the inner edge of the OHZ.}
\end{deluxetable}

\begin{figure*}
    \begin{center}
        \includegraphics[width=1.1\textwidth]{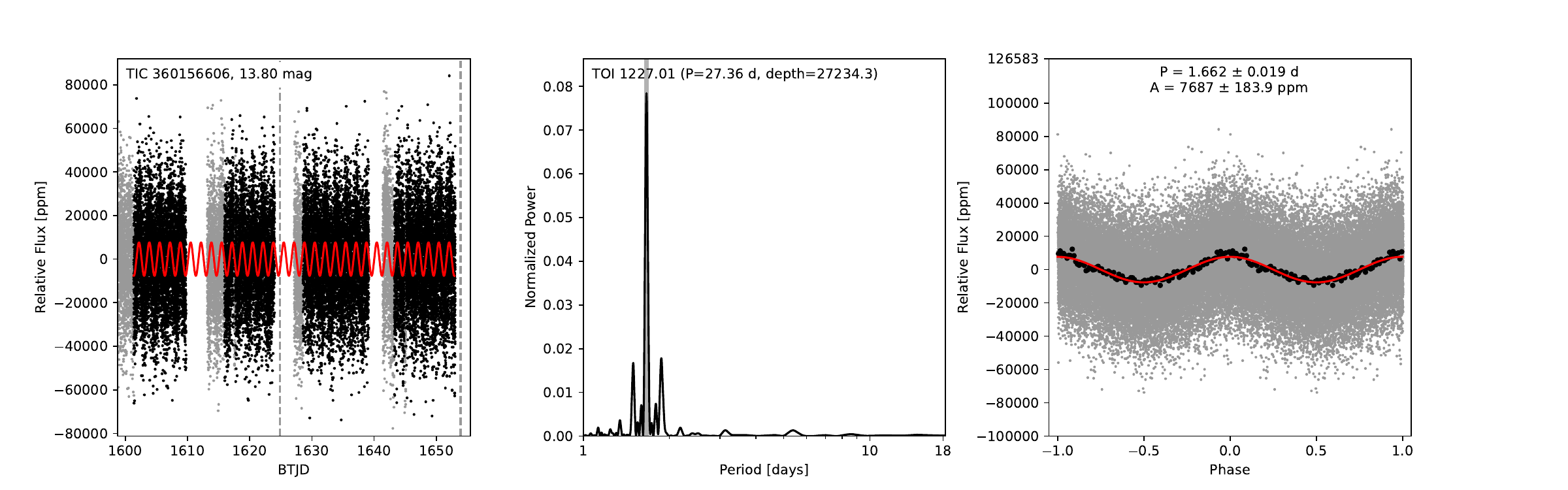} \\
        \includegraphics[width=\textwidth]{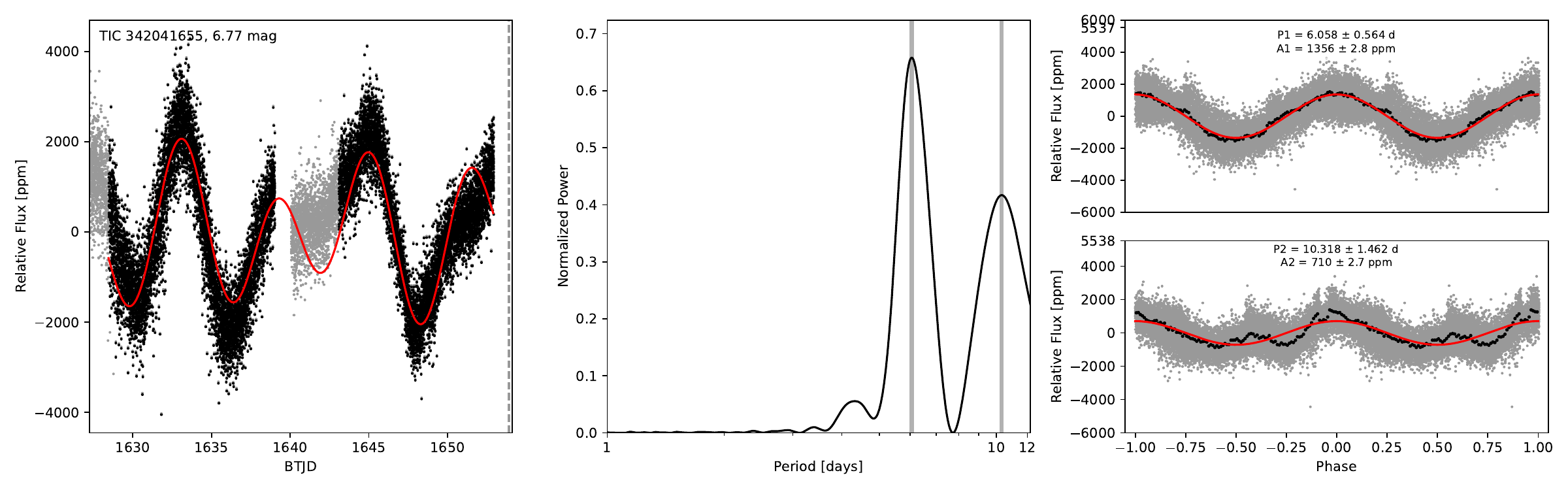} \\
        \includegraphics[width=1.1\textwidth]{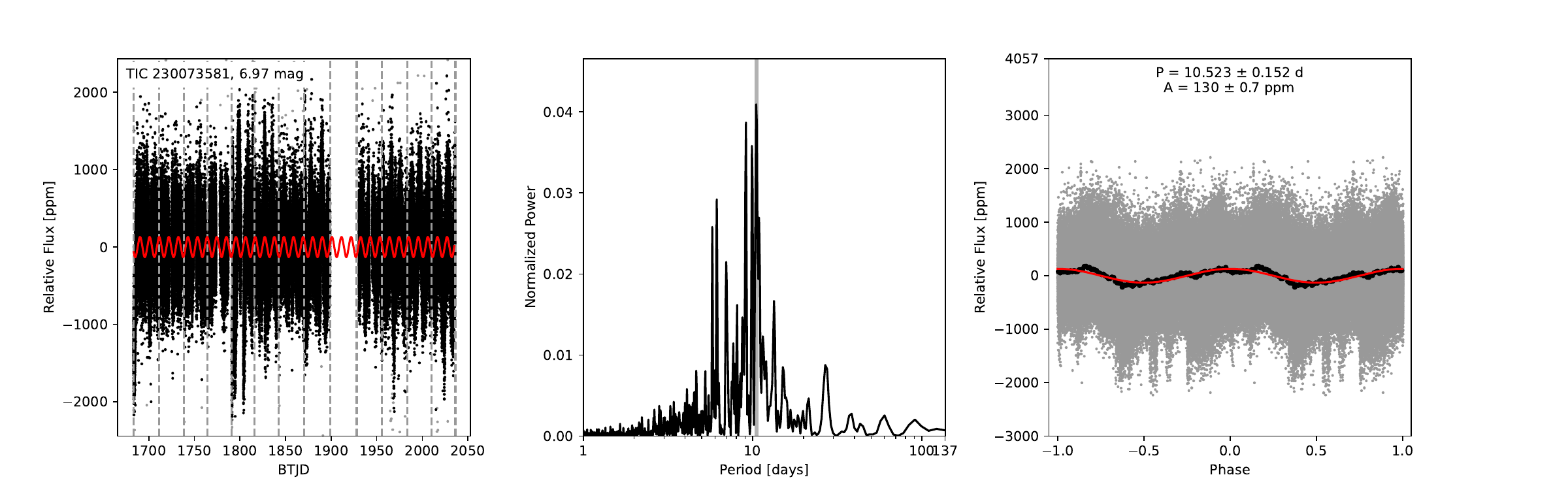}
    \end{center}
  \caption{TESS light curves (left column), Lomb-Scargle periodograms (center column), and phase-folded light curves (right column) for TOI-1227 (TIC~360156606; top row), HD~142415 (TIC~342041655; center row), and HD~147379 (TIC~230073581; bottom row). The red curves represent sinusoidal fits to the observed light curves. The dashed gray vertical lines separate individual TESS sectors, and the solid gray vertical lines indicate the selected period from the periodogram. The gray data points in the left panels are data that were not included in the variability analysis. The right panels show all TESS data in gray, and the binned data as black points.}
  \label{fig:var_selected}
\end{figure*}

In order to highlight the effects of stellar variability on HZ planets, we select TOI-1227~b, HD~142415~b, and HD~147379~b to describe in more detail. The TESS light curves (left column), Lomb-Scargle periodograms (center column), and phase-folded light curves (right column) of TOI-1227, HD~142415, and HD~147379 are shown in the top, middle, and bottom row panels of Figure~\ref{fig:var_selected}, respectively. We then more broadly discuss how stellar variability may impact exoplanet habitability in Section~\ref{sec:implications}. 

%%%%%%%%%%%%%%%%%%%%%%%%%%%%%%%%%%%%%%%%%%%%%%%%%%%%%%%%%%%%%%%%%%%%

%%%

\subsection{TOI-1227}
\label{sec:toi1227}

Of the 9 targets included in our sample, TOI-1227 has the highest amplitude of stellar variability. TOI-1227~b is also the only planet in our sample with an orbit that is consistent with being circular, which makes interpretations regarding changes in incident flux received by the planet simpler than other targets in our sample due to the absence of additional variability caused by orbital effects. TOI-1227 has an effective temperature of 3072~K and is one of only two M-dwarfs in our sample. TESS observed TOI-1227 (TIC~360156606) during sectors 11, 12, 38, 64, and 65, and TESS is anticipated to revisit the star in sectors 99 and 100. Our variability analysis utilizes sectors 11 and 12, which were observed during the TESS Prime Mission. 

This system has a single exoplanet, TOI-1227~b, that orbits at a distance of 0.0886~AU in a circular orbit near the center of the conservative HZ. While TOI-1227~b does not have a reported mass in the NASA Exoplanet Archive, it was found to have an upper mass limit of $<1.7M_J$ \citep{Mann22} with a radius of 0.854~$R_J$, making it a Jupiter-sized exoplanet. As shown in Table~\ref{tab:sample} and the top row of panels in Figure~\ref{fig:var_selected}, the host star in this system has a variability amplitude of 7687~ppm and a variability period of 1.662~days. This induces total variations in the planet's received flux of in the range of 432--438~$W/m^{2}$. The estimated changes to equilibrium temperature caused by the stellar fluctuations in flux range from 208.93~K to 209.74~K. With this star being the most extreme case of variability within our sample in terms of stellar variability amplitude, we can already see that the varying luminosity of the host star does not have a significant impact on the estimated equilibrium temperature of the planet.

%%%%%%%%%%%%%%%%%%%%%%%%%%%%%%%%%%%%%%%%%%%%%%%%%%%%%%%%%%%%%%%%%%%%

\subsection{HD 142415}
\label{sec:hd142415}

HD~142415 (TIC~342041655) is one of three stars in our sample that exhibits significant variability at two periodicities (see center row panels in Figure~\ref{fig:var_selected}), with variability amplitudes of 1356 and 711~ppm and variability periods of 6.058 and 10.318~days, respectively. HD~142415 has the second highest stellar variability amplitude in our sample after TOI-1227 and its planet has the second highest orbital eccentricity. HD~142415 is a Sun-like G star with an effective temperature of 5940~K and a mass of 1.07~$M_\odot$. HD~142415 was observed in TESS sectors 12, 39, 65, 66, and 93, and will be revisited by TESS in sectors 100--103. Our variability analysis uses sector 12 from the TESS Prime Mission. 

\begin{figure}
    \begin{center}
        \includegraphics[width=\linewidth]{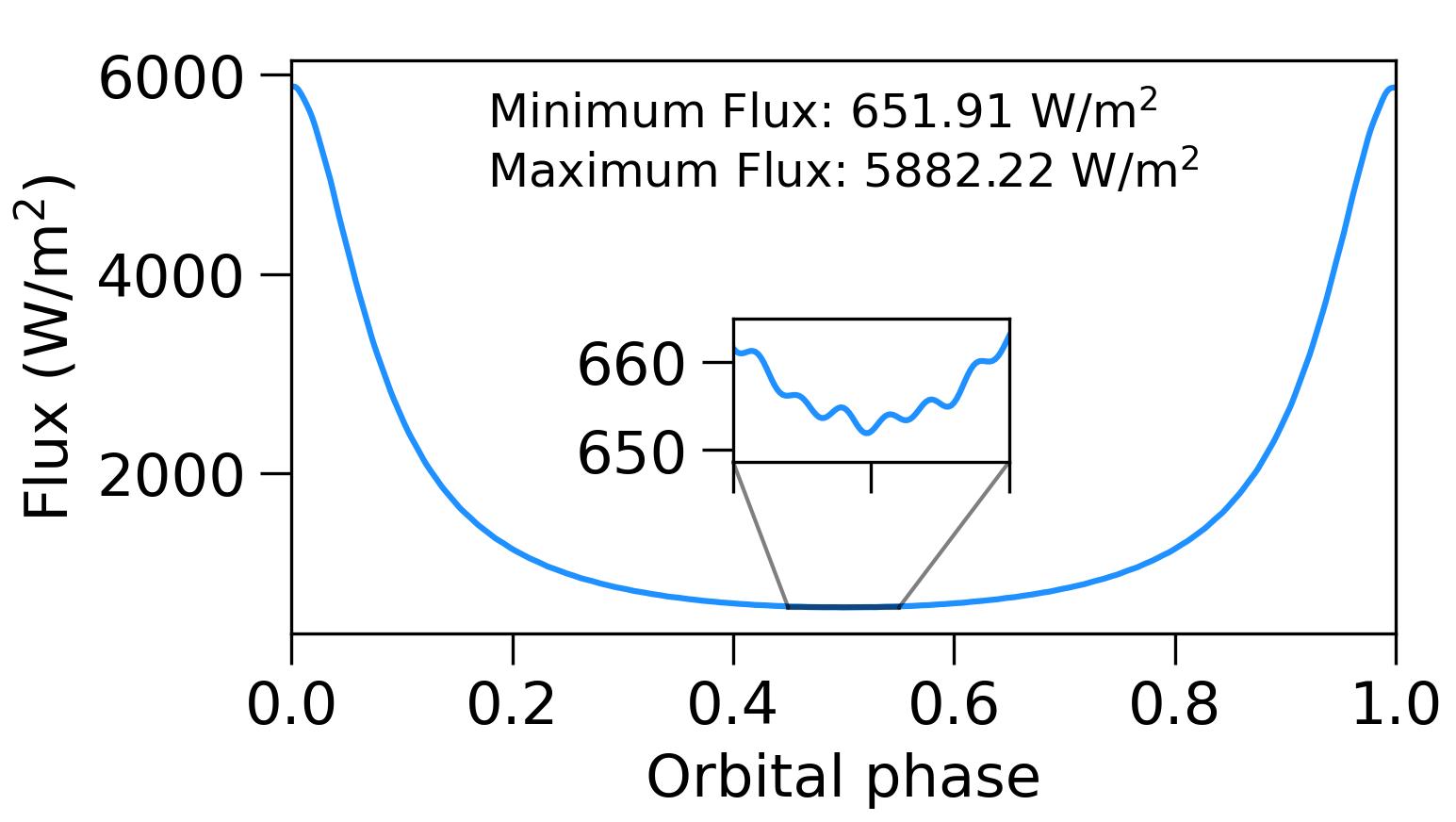}
    \end{center}
  \caption{Flux received by HD~142415~b from combined orbital and stellar variability effects over the course of one orbital period. The changes in flux are primarily dominated by the effects of the planet's eccentricity, but the additional fluctuations in flux caused by the host star's double-sinusoidal variability (see center row panels of Figure~\ref{fig:var_selected}) can be seen in the zoomed-in sub-panel.}
  \label{fig:combined_flux}
\end{figure}

The only known exoplanet in this system is HD~142415~b, which is a giant planet with a measured $m_p\sin{i}$ of 1.67~$M_J$, an orbital period of 386.3~days, and an eccentricity of 0.5. While the star's flux due to its variability only changes by 2\%, the flux received at the top of the planet's atmosphere can range from 653--5890~$W/m^{2}$---largely due to the planet's orbital eccentricity. In Figure~\ref{fig:combined_flux}, we show how the flux variations caused by the double-sinusoidal stellar variability compare to the flux variations caused by the planet's orbital eccentricity. When considering the equilibrium temperature changes attributed to only the stellar variability, we find that the equilibrium temperature only ranges from 401.02~K to 401.44~K. Clearly, in this scenario the planet's orbital eccentricity dominates changes in received flux and equilibrium temperature compared to the double-sinusoidal nature of the star's variability.

%%%%%%%%%%%%%%%%%%%%%%%%%%%%%%%%%%%%%%%%%%%%%%%%%%%%%%%%%%%%%%%%%%%%

\subsection{HD 147379}
\label{sec:hd147379}

HD~147379 (TIC~230073581) has the second-lowest amplitude of stellar variability and has the smallest mass planet among the 9 targets we have explored, but it also has the longest baseline of observations with TESS. HD~147379 is located in the northern continuous viewing zone of TESS, resulting in extensive observations in sectors 14--26, 40--41, 47--60, and 73--86. The long baseline allows for an extended periodogram search range, but our variability analysis on the TESS Prime Mission data (sectors 14--26) only showed a signal 130~ppm at a variability period of 10.523~days (see bottom row panels of Figure~\ref{fig:var_selected}). 

The star in the HD~147379 system is a warmer M-dwarf with an effective temperature of 4090~K and a mass of 0.58~$M_\odot$. The system hosts a single exoplanet, HD~147379~b, which is a sub-Saturn with a mass of 0.068~$M_J$ and is the smallest mass planet in our sample. Additionally, the planet orbits its host star in a nearly circular orbit within the inner edge of the OHZ with a semimajor axis of 0.323~AU and an eccentricity of 0.063. 

Despite its close proximity to the M-dwarf host, HD~147379~b experiences very little effect from the variability of its host star. Similar to the case of HD~142415~b, nearly all of the variation seen with the flux is due to the eccentricity of the planet, with the changes in flux ranging 1230--1584~$W/m^2$. The orbital effects caused by the planet's eccentricity once again dominate over any effects caused by the variability of the star, and consequently, the low amplitude of the host star's variability has little impact on the planet's climate. This becomes more apparent when investigating the effect of the host star's variability on the equilibrium temperature of the planet, which only ranges from 289.08~K to 289.10~K. For this system, and many others among the 9 targets we have studied in this work, the variability of the host star is shown to have an insignificant effect of the climate of the planet it hosts.

%%%%%%%%%%%%%%%%%%%%%%%%%%%%%%%%%%%%%%%%%%%%%%%%%%%%%%%%%%%%%%%%%%%%

\subsection{Implications of Stellar Variability on Planetary Habitability}
\label{sec:implications}

For the planets in our sample, stellar variability overall had an insignificant impact on the estimated equilibrium temperatures of the planets when compared to the changes in temperature caused by their orbital eccentricities. This is an important distinction since the orbital eccentricity has the potential to greatly influence planetary climate \citep{williams2002,dressing2010,kane2012e,barnes2013a,bolmont2016a,way2017a}, and can indeed increase water loss rates \citep{kane2020e,palubski2020a}. The most extreme change in equilibrium temperature is more than 200~K for HD~238914~b (see Table~\ref{tab:teq}), which is the highest eccentricity planet in our sample ($e=0.56$). The most extreme equilibrium temperature change due to stellar variability, on the other had, is only 1~K for TOI-1227~b. Even the planet with the lowest non-zero eccentricity in our sample, HD~137379~b, experiences a 17~K temperature change. Therefore, the host star's variability is unlikely to have a significant impact on the potential habitability of the HZ exoplanets in our sample.

However, the stars in our sample have relatively low variability amplitudes when compared to the larger sample of variable stars in the TESS-SVC \citep{fetherolf2023b}. The average variability amplitude of the stars in the TESS-SVC is 11,350~ppm, which is 1.5 times higher than the variability amplitude of TOI-1227, and 11\% of all variable stars have variability amplitudes higher than that of TOI-1227. Therefore, HZ planets around more extreme variable stars---which are not represented in our sample---may experience more significant impacts on their climates. When investigating the properties of known exoplanets around variable host stars, \citet{simpson2023} emphasized that there is an observational bias against discovering exoplanets around variable stars---especially when it comes to smaller exoplanets. This is largely because the small signals of exoplanets are very difficult to extract from the much larger signals of noisy variable stars. Stellar variability has also been attributed to the discovery of false positive exoplanets \citep{Robertson14,Robertson14-1,Robertson15,sullivan2015,kane2016a,Hojjatpanah20,Prajwal22,simpson2022a} and makes it more difficult to measure accurate host star properties \citep{Ribas06, Chabrier07, Lopez-Morales07, Kraus11}, which often has led to underestimated planet radii \citep{Seager07, ciardi2015a, hirsch2017, kane2014a, kane2018a}. Furthermore, stellar binarity can lead to incorrect estimates of planet radii and habitable zone boundaries if the planet is not around the primary star, as is typically assumed \citep[][Burns-Watson et al., submitted]{Gaidos16}.  

\begin{figure*}
    \begin{center}
        \includegraphics[width=18cm]{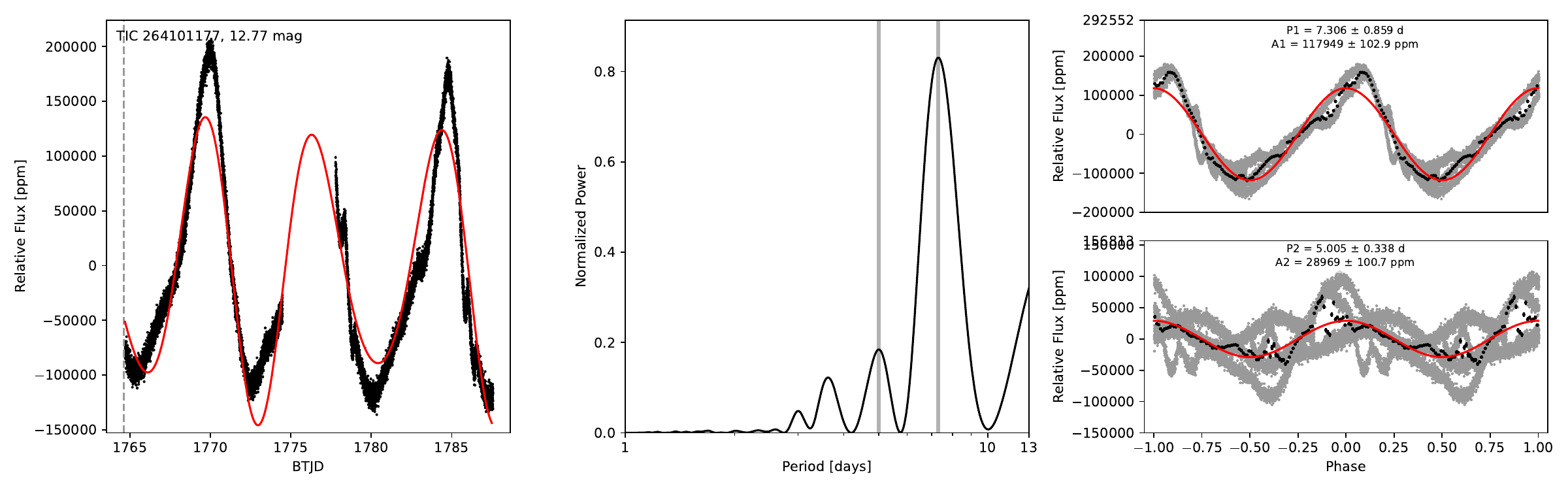}
    \end{center}
  \caption{Same as Figure~\ref{fig:var_selected}, but for TIC~264101177, which exhibits extremely high amplitude variations. This type of variable star is not represented in our sample of HZ planets around variable stars---likely due to observational bias.}
  \label{fig:extreme_var}
\end{figure*}

Regardless of the current observational bias, exoplanets certainly exist around variable host stars. However, the outstanding question that we are trying to address in this work is how stellar variability may impact the climates of HZ exoplanets. To demonstrate an extreme example of stellar variability, we show the TESS light curve, periodogram, and phase folded light curve of TIC~264101177 in Figure~\ref{fig:extreme_var}, which has a variability amplitude of 117,949~ppm. If this star were to have an exoplanet, its signal would certainly be hidden in the extreme nature of the variability amplitude, thus limiting the discovery of its potential planet(s). However, the climate of a potential planet around TIC~264101177 would more likely be affected by the host star's extreme variability compared to the 9 planets in our sample. Since this type of star is not present in our sample of HZ exoplanets around variable stars, we turn to investigating the effects of stellar variability through a simulated planetary system in Section~\ref{sec:vplanet}. In addition to investigating the short term effects of stellar variability on a HZ exoplanet's climate, we will also be able to investigate how stellar variability could impact a planet's climate over stellar evolutionary timescales. 

%%%%%%%%%%%%%%%%%%%%%%%%%%%%%%%%%%%%%%%%%%%%%%%%%%%%%%%%%%%%%%%%%%%%

\section{Climate Effects of Variable Host Stars}
\label{sec:vplanet}

%%%%%%%%%%%%%%%%%%%%%%%%%%%%%%%%%%%%%%%%%%%%%%%%%%%%%%%%%%%%%%%%%%%%

\subsection{VPLanet Simulations}

\begin{figure*}
    \begin{center}
        \includegraphics[width=1.05\linewidth]{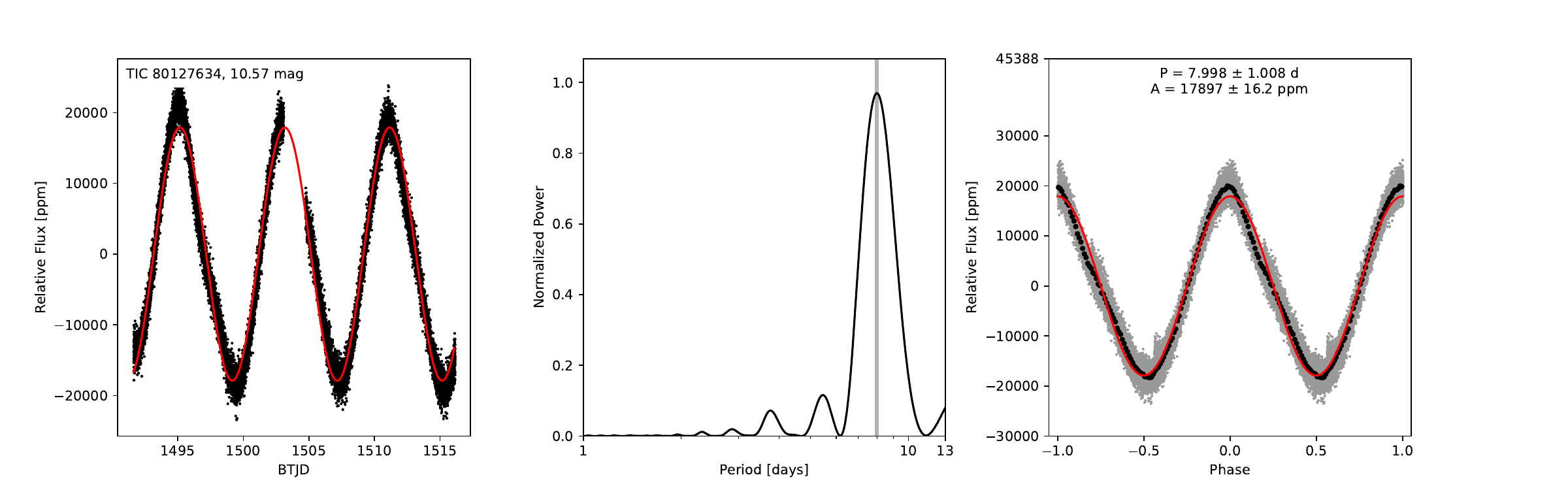}
        \includegraphics[width=7.0cm]{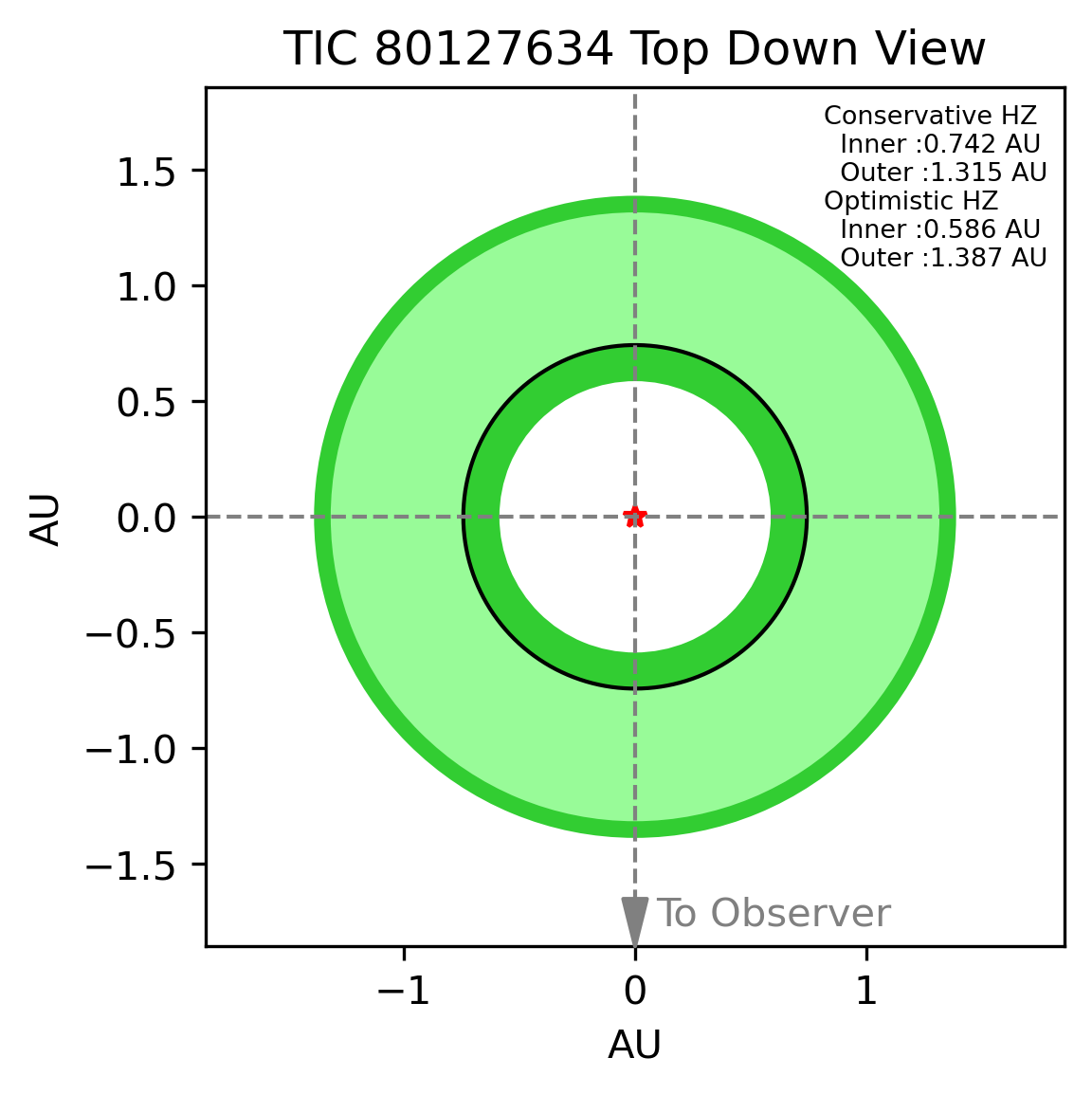}
        \includegraphics[width=10.0cm]{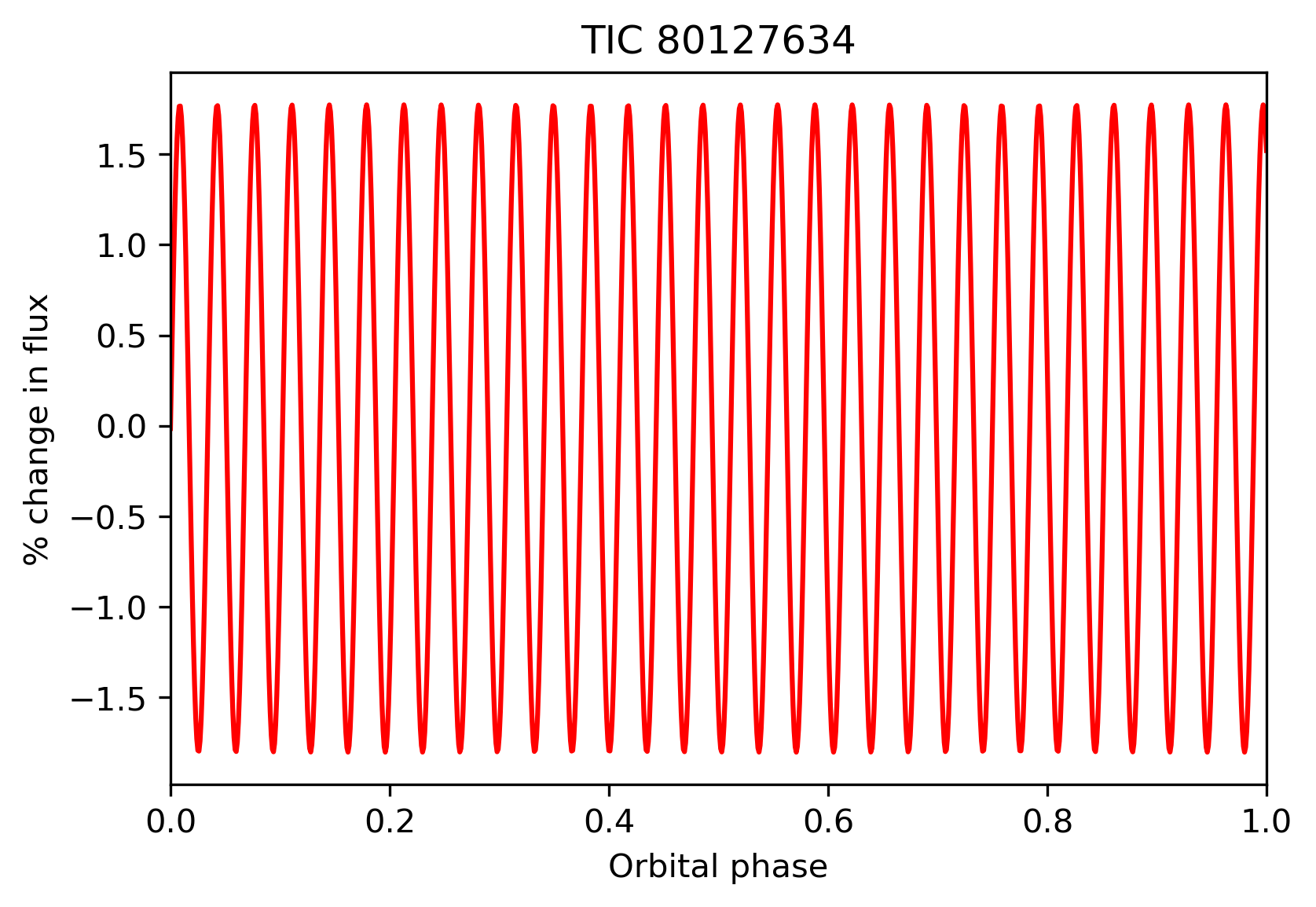}
    \end{center}
  \caption{Top: Same as Figure~\ref{fig:var_selected}, but for TIC~80127634, which is a G star that exhibits high amplitude variations that is used for the water loss simulations. Left: Orbit architecture of a hypothetical Earth-like planet in a circular orbit within the HZ of TIC~80127634. The black circle shows the simulated planet orbit. The light shaded region shows the CHZ and the dark shaded region shows the larger bounds of the OHZ. Right: Percent change in the incident flux received by the hypothetical Earth-like planet.}
  \label{fig:hypo}
\end{figure*}

Given that stellar flux is the primary contributor to a planet's energy budget, having a variable host star may impact the climate, seasons, and evolution of terrestrial exoplanets. Gaining a better understanding of how planets are affected by variable host stars will improve our ability to estimate the habitability of planets in similar star systems. For this analysis we focus on the variable star TIC~80127634, which is a G-type star with a mass of 0.99~$M_\odot$ and radius of 0.82~$R_\odot$ that was observed in sector 7 of TESS observations (see top panels of Figure~\ref{fig:hypo}). This star shows a single sinusoidal pattern of photometric variability with an amplitude of 17,897~ppm, which is more than twice as large as the highest amplitude variable star in our sample (see Table~\ref{sec:sample}), and a variability period of 7.99~days. This star was selected to better represent highly variable stars, which are not represented in our sample of 9 HZ exoplanets around variable stars. Using this star will further inform how exoplanet climates may be affected by variable host stars.

Since TIC~80127634 is not currently known to host any planets, we defined a hypothetical system where an Earth-analog orbits TIC~80127634 on the inner edge of the conservative HZ \citep[CHZ;][]{kopparapu2014,kasting1993a,kopparapu2017} boundary, which is defined to be an orbit with a semi-major axis of of 0.742~AU and period of 234.63~days. We chose to place the planet on the inner edge of the HZ such that the star's variability has a more extreme effect on its hypothetical planet, and to better gauge how stellar variability can impact water loss on an Earth-like planet. The architecture of this hypothetical system is shown on the bottom left panel of Figure~\ref{fig:hypo}, where a top-down view of the host star and planet can be observed in addition to the boundaries of the conservative (light green) and optimistic (dark green) HZs of the system. We have chosen to place the planet in a circular orbit in order to isolate changes in the flux it receives from its variable host from fluctuations caused by orbital effects due to eccentricity. 

Shown in the bottom right panel of Figure~\ref{fig:hypo}, we show the percent change in flux for TIC~80127634 based on an average value of 1480~W/m$^2$. The range of flux received by the hypothetical planet and the corresponding range in equilibrium temperature is reported in Table~\ref{tab:teq}. For this star, the flux ranges from a minimum flux of 1452~W/m$^2$ to a maximum flux of 1505~W/m$^2$, which corresponds to approximately a 4\% total change in stellar flux. For the hypothetical planet within the system, these variations cause changes in the planet's equilibrium temperature ranging from a minimum of 453.93~K to a maximum of 458.01~K, or a total change of nearly 4~K every 7.99~days. 

We used the aforementioned orbital and stellar parameters as inputs for \texttt{VPLanet} \citep{barnes2020} to evaluate the difference in water loss which occurs on an Earth-like planet with both a variable and non-variable host star. The non-variable host star will hereafter be referred to as a quiet star. VPLanet is a publicly-available software package that is capable of simulating various planetary processes over geological timescales. For this study, we used the \texttt{AtmEsc} and \texttt{STELLAR} modules to simulate energy-limited and diffusion-limited atmospheric escape for Earth-like planets due to the star's evolving XUV luminosity. \texttt{AtmEsc} is able to predict the loss of surface water on a planet by calculating the escape of hydrogen to space.

\begin{figure*}
    \begin{center}
        \includegraphics[width=0.95\textwidth]{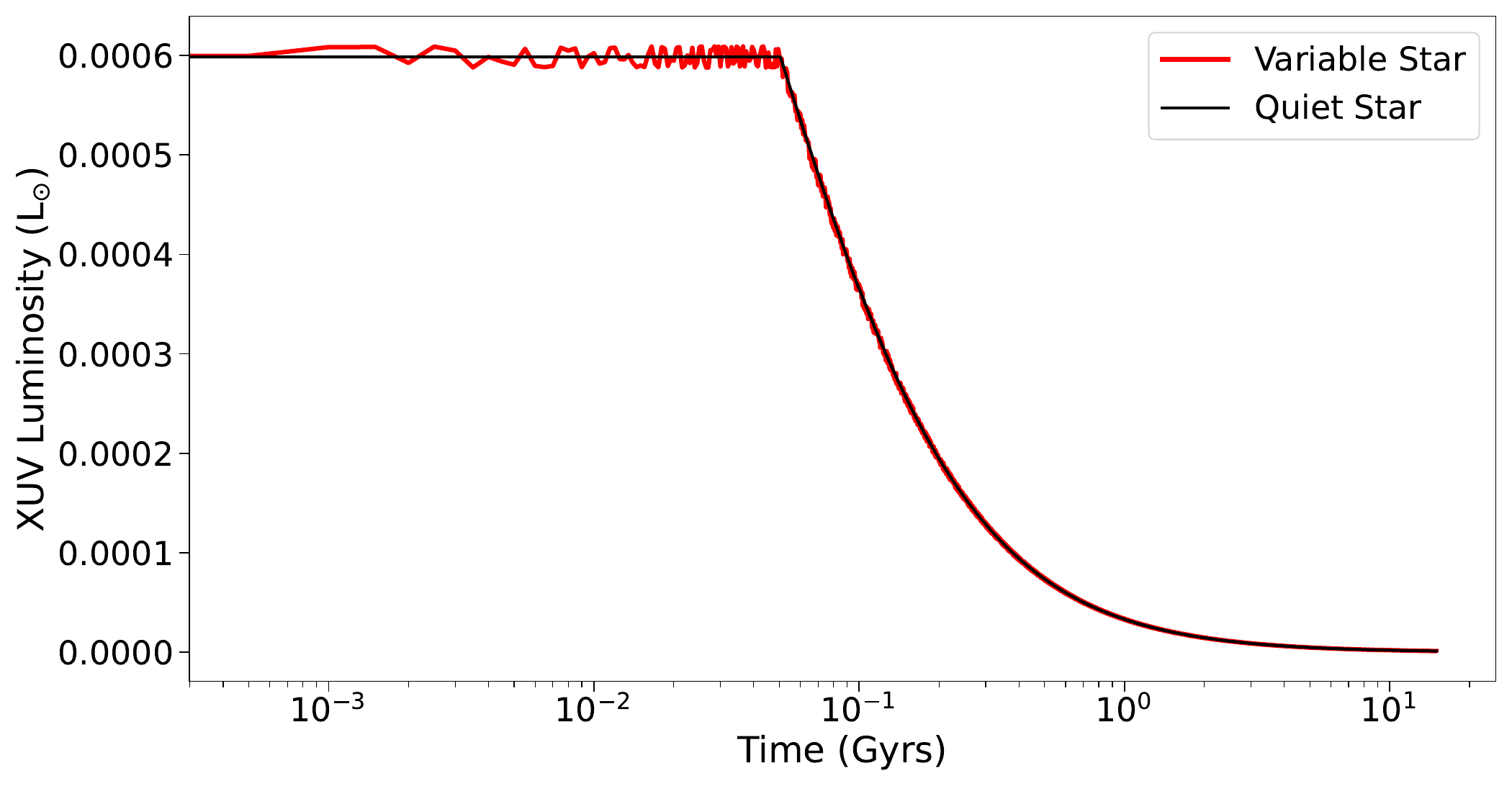}
    \end{center}
    \caption{A comparison of the XUV luminosity evolution for the quiet and variable stars used in the water loss simulations. The variable star has fluctuations in XUV luminosity, but generally follows the same curve as the quiet star. The fluctuations are less visible after the initial saturation phase in the figure, but they do persist throughout the entire simulation.}
    \label{fig:xuvcompare}
\end{figure*}

We added new functionality to \texttt{VPLanet}'s \texttt{STELLAR} module to simulate XUV variability, which we assume is a constant fraction of the bolometric luminosity. We implemented a simple sinusoidal model in which the bolometric luminosity oscillates with a constant amplitude and frequency, which can be set to the values for known stars. We used the constant luminosity option included in \texttt{VPLanet} to represent the quiet star, and directly compared mass loss rates between the two models to quantify how stellar variability affects water loss rates on habitable planets. The primary factor that determines the water loss rates predicted by \texttt{AtmEsc} is the extreme ultra-violet (XUV) luminosity of the host star, such that it is important to ensure that the average XUV luminosity of the variable star closely matches the XUV luminosity of the quiet star to accurately model the water loss rates. In \texttt{AtmEsc}, the initial XUV luminosity is estimated to be a user-defined fraction of a star's total luminosity, which we defined to be $10^{-3}$\,L$_{\odot}$. The XUV evolution of the star throughout the simulation follows the \citet{ribas2005evolution} model, which includes an initial saturation phase followed by a power-law decrease in XUV luminosity. A more detailed description of this power law can be found in Equation 146 of \citet{barnes2020}. Figure \ref{fig:xuvcompare} illustrates the XUV evolution for both the variable and quiet star used in our models. The saturation phase occurs for the first 0.1\,Gyrs of the simulation, and then the XUV luminosity follows the aforementioned power law. It can be seen that the variable star has fluctuating XUV luminosity whereas the quiet star directly follows the \citet{ribas2005evolution} model, but the average variable star XUV luminosity closely follows the quiet star XUV luminosity model for the full duration of the simulation.

We chose to test cases where the planet starts with a surface liquid water abundance that is 1, 5, 10, and 50 times greater than Earth's oceans, hereafter referred to as terrestrial oceans (TO), for both of a quiet and variable host star. In total, we ran 10 \texttt{VPLanet} simulations in this work. For all simulations the orbital configuration, stellar mass, stellar radius, average stellar luminosity, planet radius, planet mass, and planet's initial atmospheric composition remained constant. The stellar radius and mass for both stars were defined to be the same as TIC~80127634, while the planet's mass, radius, and initial atmospheric composition was set to be the same as Earth's. The average luminosity for both stars was defined to be 0.598\,L$_{\odot}$, with the variable star luminosity fluctuating between 0.588--0.608\,L$_{\odot}$ with a periodicity of 7.99\,days. To simulate a variable star with \texttt{AtmEsc}, we used these values as inputs into the new ``\texttt{sinewave}'' stellar model that mimics the sinusoidal nature of variable stellar flux. The initial age of both stars were defined to be 50\,Myr years and the simulations were set to run for as long as 15\,Gyr, or until all the surface water was removed from the planet. The iterative steps of the simulations were every 50,000\,years.

%%%%%%%%%%%%%%%%%%%%%%%%%%%%%%%%%%%%%%%%%%%%%%%%%%%%%%%%%%%%%%%%%%%%

\subsection{Water Loss Results}
\label{sec:waterloss}

\begin{figure*}
    \begin{center}
        \includegraphics[width=0.95\textwidth]{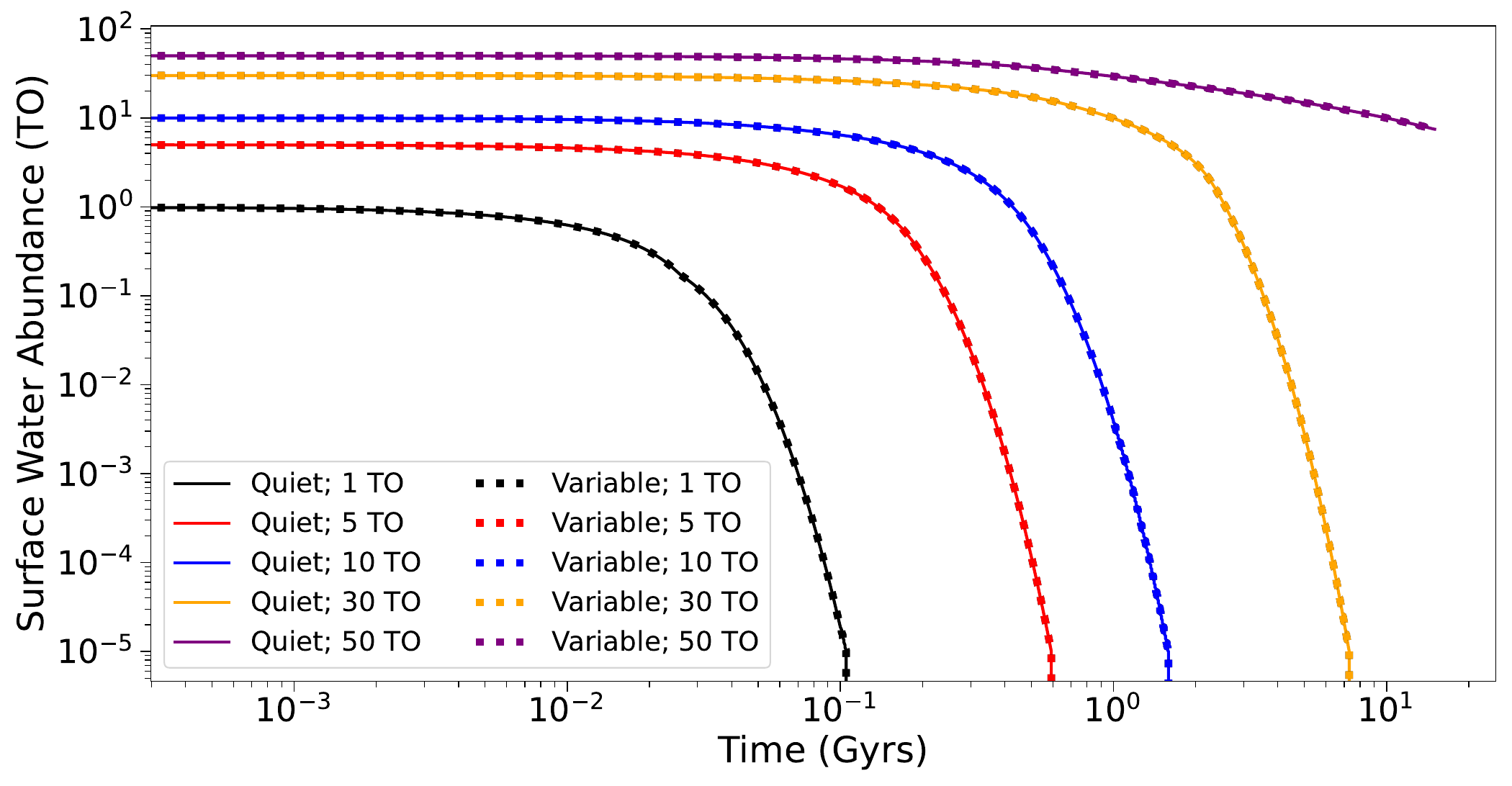}
    \end{center}
  \caption{Water loss over time for a hypothetical Earth with varying amounts of surface water on the inner edge of the CHZ. The simulations with the quiet host star and variable host star are show in in the solid and dotted lines, respectively. Despite the fluctuations in luminosity (see Figure~\ref{fig:xuvcompare}), the variable host star shows identical water loss rates to the quiet star for every TO case. Nearly all simulations lost all water before the end of the simulation, expect the 50\,TO case which had 7.47\,TO of water remaining after 15\,Gyrs of simulation time.}
  \label{fig:waterloss}
\end{figure*}

Figure~\ref{fig:waterloss} demonstrates the change in surface water abundance as a function of time for each of the 10 simulations. All simulations are similar in that they experience little-to-no water loss for a period of time near the beginning of their simulations. It can be seen that both the variable and quiet host stars cause identical water loss rates for a given initial water inventory. The time needed for both star types to desiccate their planet of water was identical as well, with the 1\,TO, 5\,TO, 10\,TO, and 30\,TO cases requiring 0.105\,Gyr, 0.592\,Gyr, 1.590\,Gyr, and 7.300\,Gyr, respectively. The 50\,TO case did not completely deplete its water inventory after 15\,Gyrs, with 7.47\,TO remaining on the planet by the end of the simulation.

Overall, our results are indicate that variable stars have a negligible impact on the water-loss rates of Earth-like planets. This finding is within our expectations given the sinusoidal nature of the stellar model used for the variable star, which provides a balance between higher and lower luminosities that could impact the water inventory. However, for simplicity in this work, we have only tested a single variable star with only variations in the the initial water inventory of the planet. This work needs to be further developed through further exploration of the variable star parameter space, such as testing how different variability periods, variability amplitudes, and stellar spectral types could impact the water-loss rates of Earth-like planets. In particular, testing long-period stellar variations or asymmetric fluctuations (opposed to sinusoidal) may result in water-loss rates that differ from quiet stars. Since the water loss rates predicted by \texttt{AtmEsc} are solely dependent on a star's XUV flux, it may be valuable to also revisit this study with other stellar models that incorporate additional atmospheric escape processes. Further examining how stellar activity may affect atmospheric spectroscopy of exoplanets, as was discussed in \citet{knutson2011spitzer}, might be worth investigating in future works as well. Specifically, how the magnitude of host star variability can affect the transit depth of spectral features in exoplanet atmospheres, and whether this has similar spectral effects to that of changes in cloud coverage or atmospheric composition. Ultimately, this study provides insight into how the climates of rocky worlds may be impacted by variable host stars, but further investigation is required to validate whether these findings are consistent across a variety of variable stars.

%%%%%%%%%%%%%%%%%%%%%%%%%%%%%%%%%%%%%%%%%%%%%%%%%%%%%%%%%%%%%%%%%%%%

\section{Implications for Exomoons}
\label{sec:exomoons}

Nearly all of the cases that we have explored in this study are of variable host stars and giant planets in the HZ (see Section~\ref{sec:sample_stats}). Despite their location within the HZ of these stars and the minimal effect their host star's variability has on their climates, giant planets are unlikely targets for future missions aiming to characterize potentially habitable exoplanet atmospheres due to the nature of their composition. However, it is worth discussing the potential scenario where these giant planets may host moons that could maintain long-term dynamical stability \citep{lammer2014b,kane2017c,harada2023}. In contrast to their parent planets, exomoons are notable candidates for habitability beyond Earth due to the equilibrium temperatures that can be achieved from the heat redistribution efficiency from the planet \citep{forgan2014b,heller2014c}.

There are various sources of energy for moons, such as stellar flux, thermal emission and reflected light from the planet, and tidal heating, many of which have a time variable component \citep{heller2012,heller2013a,hinkel2013b,dobos2017}. Indeed, the combination of these energy sources can potentially drive the climate of an exomoon into a post runaway greenhouse state \citep{heller2015b}, producing conditions more similar to Venus than the temperate conditions of Earth \citep{kane2014e,way2020,kane2024b}. For the majority of cases studied in this work, the amplitude of the incident flux variations resulting from stellar variability are considerably less than those from the other factors, such as orbital eccentricity.  For example, Figure~\ref{fig:combined_flux} shows that the incident flux effects on HD~142415~b from the variability of the host star are minor compared those resulting from the eccentricity of the planetary orbit. For our sample of variable stars (see Table~\ref{tab:sample}), the variability amplitudes are typically on the order of several hundred ppm, although their modest amplitudes are likely a result of an observational bias against detecting exoplanets around variable stars (see Section~\ref{sec:implications}). A significant outlier within our sample is the case of TOI-1227, with a variability amplitude that approaches $\sim$10,000~ppm, which then becomes comparable to the incident flux variations that result from the orbital motion of a moon around a giant planet \citep{hinkel2013b}. The hypothetical case of a planet in the HZ of TIC~80127634, shown in Figure~\ref{fig:hypo}, presents a more extreme scenario where the variability of the host star can dominate the incident flux variations received by the planet, along with any potential exomoon orbiting the planet. Thus, stellar variability could indeed be a major component in driving the climate evolution of an exomoon, and may even be more common that previously suspected once the observational bias against detecting exoplanets around variable stars is accounted for.

%%%%%%%%%%%%%%%%%%%%%%%%%%%%%%%%%%%%%%%%%%%%%%%%%%%%%%%%%%%%%%%%%%%%

\section{Conclusions}
\label{sec:conclusions}

To better understand the effect of stellar variability on a planet's climate, we analyzed a selection of 9 variable star targets that are known hosts to at least one exoplanet with an orbit that spends more than 50\% of its time in the OHZ. We performed an updated variability analysis on these stars using all data available from the TESS Prime Mission (sectors 1--26) following a similar procedure to \citet{fetherolf2023b}. The 9 stars in our sample have variability amplitudes of 124--7687~ppm and variability periods of 1.662--10.725~days. The effective temperature of the star, variability amplitude and period, and planetary orbital period and eccentricity are listed for each target in our sample in Table~\ref{tab:sample}.

We calculated the estimated top-of-atmosphere insolation flux received by the planets and their expected equilibrium temperatures. Due to significant differences caused by eccentricity and stellar variability, the fluxes and equilibrium temperatures are separately calculated and reported in Table~\ref{tab:teq}. Only one of our targets (TOI-1227~b) had an eccentricity equal to zero, such that the fluctuations in insolation flux were due to the variability of the host star alone. In cases where the planet held an eccentric orbit around its variable host, the orbital effects dominated over the effects due to stellar variability. The relative impact of stellar variability compared to the eccentricity on the flux received by the planet can especially be observed in Figure~\ref{fig:combined_flux}. In our sample, TOI-1227 experienced the largest changes in equilibrium temperature caused by just stellar variability due to it having the largest stellar variability amplitude, but these changes only amounted to a change of 1~K. Overall, for the targets in our sample, we find that stellar variability has a negligible effect on the equilibrium temperature of planets in the HZ. 

At the same time, we also show that there is an observational bias against observing exoplanets around stars with extreme variability. We find that 11\% of all variable stars have variability amplitudes larger than highest variability amplitude star in our sample (TOI-1227). Therefore, we investigate the scenario of a hypothetical planet around a more extreme variable star in Section~\ref{sec:vplanet}. The variable star in this scenario, TIC~80127634, is a variable star reported in the TESS-SVC with an amplitude of 17,897~ppm and a period of 7.99~days. We compare this scenario to that of TOI-1227, as each of their architectures involve a planet with a circular orbit that allows this study to isolate the effects of stellar variability. From the values in Table \ref{tab:teq}, we observe a total change in equilibrium temperature of 4.08 K for the hypothetical Earth, while only 1.09~K for TOI-1227. 

The results of VPLanet simulations using quiet and variable host stars provided evidence that stellar variability has a negligible effect on the rate of water loss for Earth-like planets. As shown in Figure \ref{fig:waterloss}, the water loss rates for both the quiet and variable stars were identical for the entirety of their simulations, regardless of the initial water inventory of the planet. Furthermore, the time of complete water loss was the same for both the quiet and variable host star. It is possible that other variable star types, such as those with longer periods of variability, may result in water loss rates which differ from quiet stars, but this would need to be confirmed in future studies.

This work is only a small step towards understanding the relationship between variable stars, planetary properties, and their climates. Our comparison of flux variations due to stellar variability and orbital eccentricity generally assume that the orbital period is much greater than that of the stellar variability. However, resonances between these two effects, particularly for relatively short period planets in the HZ of M dwarfs, may produced combined effects that either expand or decrease the width of the HZ. As discussed in Section~\ref{sec:implications}, the variable stars explored in this study have amplitudes that are below average in terms of the stars observed in the TESS-SVC. Furthermore, there is a small pool of variable star candidates that are known to host exoplanets due to observational biases against finding exoplanets around variable stars. Additional observations of variable stars and the discovery and characterization of their planets will further enable our ability to understand how planetary climates respond to their variable host stars. 

%%%%%%%%%%%%%%%%%%%%%%%%%%%%%%%%%%%%%%%%%%%%%%%%%%%%%%%%%%%%%%%%%%%%

\section*{Acknowledgements}

The authors acknowledge support from NASA grant 80NSSC24K0227 and 80NSSC24K1491 funded through the TESS Guest Investigator Program and the Astrophysics Data Analysis Program. T.F. acknowledges support from an appointment through the NASA Postdoctoral Program at the NASA Astrobiology Center, administered by Oak Ridge Associated Universities under contract with NASA. This research has made use of the Habitable Zone Gallery at hzgallery.org. The results reported herein benefited from collaborations and/or information exchange within NASA's Nexus for Exoplanet System Science (NExSS) research coordination network sponsored by NASA's Science Mission Directorate. VPLanet development has been supported by NASA grants NNA13AA93A, NNX15AN35G, 80NSSC17K048, 13-13NAI7\_0024, and 80NSSC20K0229. We also acknowledge support from the University of Washington and the Carnegie Institute for Science. This research has made use of the NASA Exoplanet Archive (\dataset[DOI: 10.26133/NEA12]{http://dx.doi.org/10.26133/NEA12}), which is operated by the California Institute of Technology, under contract with the National Aeronautics and Space Administration under the Exoplanet Exploration Program. We acknowledge the use of public TESS data from pipelines at the TESS Science Office and at the TESS Science Processing Operations Center. Resources supporting this work were provided by the NASA High-End Computing (HEC) Program through the NASA Advanced Supercomputing (NAS) Division at Ames Research Center for the production of the SPOC data products. All of the data presented in this paper were obtained from the Mikulski Archive for Space Telescopes (MAST). The specific observations analyzed can be accessed via \dataset[DOI: 10.17909/t9-nmc8-f686]{http://dx.doi.org/10.17909/t9-nmc8-f686}. STScI is operated by the Association of Universities for Research in Astronomy, Inc., under NASA contract NAS5-26555. Support for MAST for non-HST data is provided by the NASA Office of Space Science via grant NNX13AC07G and by other grants and contracts. This paper includes data collected with the TESS mission, obtained from the MAST data archive at the Space Telescope Science Institute (STScI). Funding for the TESS mission is provided by the NASA Explorer Program. STScI is operated by the Association of Universities for Research in Astronomy, Inc., under NASA contract NAS 5–26555. This research made use of Lightkurve, a Python package for Kepler and TESS data analysis \citep{Lightkurve_Collaboration18}.

%%%%%%%%%%%%%%%%%%%%%%%%%%%%%%%%%%%%%%%%%%%%%%%%%%%%%%%%%%%%%%%%%%%%

\facilities{TESS}

\software{VPlanet \citep{barnes2020},
          Astropy \citep{Astropy_Collaboration13, Astropy_Collaboration18},
          Astroquery \citep{Ginsburg19},
          Lightkurve \citep{Lightkurve_Collaboration18},
          Matplotlib \citep{Hunter07},
          NumPy \citep{Harris20}, 
          SciPy \citep{Virtanen20}}

%%%%%%%%%%%%%%%%%%%%%%%%%%%%%%%%%%%%%%%%%%%%%%%%%%%%%%%%%%%%%%%%%%%%

\bibliographystyle{aasjournal}
\bibliography{references,newrefs}

\end{document}